\documentclass[aps,prl,reprint,groupedaddress]{revtex4-2}

\usepackage{bm,amsmath,mathtools,amsfonts}
\usepackage{siunitx}
\usepackage{graphicx, color}

\begin{document}


\title{Self-Trapping of Microorganisms Steering Toward their Own Trail}


\author{Aymeric Lutier}
\affiliation{Laboratoire Matière et Systèmes Complexes (MSC), Université Paris Cité \& CNRS, 75013 Paris, France}
\author{Frédéric van Wijland}
\affiliation{Laboratoire Matière et Systèmes Complexes (MSC), Université Paris Cité \& CNRS, 75013 Paris, France}
\affiliation{Yukawa Institute for Theoretical Physics, Kyoto University, Kyoto, 606-8502, Japan}

\author{Jean-Baptiste Fournier}
\affiliation{Laboratoire Matière et Systèmes Complexes (MSC), Université Paris Cité \& CNRS, 75013 Paris, France}


\date{\today}

\begin{abstract}
Active matter systems comprise self-propelled particles that move on a substrate while leaving chemical trails that influence other particles through chemotaxis (e.g., slime-depositing bacteria). Orientational chemotaxis manifests as a torque that steers the particle toward the chemical gradient. As each particle is coupled to its own trail, the dynamics exhibits an instability: when the particle gently diffuses, it abruptly transitions to trajectories with a radius of curvature comparable to its own size, becoming apparently trapped. We argue that, contrary to intuition, this trajectory instability occurs for any chemotactic coupling strength. Depending on the coupling regime, this arises either through a potential-barrier first-passage problem or from a rare event analysis.
\end{abstract}

\maketitle

Active matter encompasses systems of particles that extract energy from their surroundings and convert it into motion or internal state changes. These systems cover a broad diversity of types---from cytoskeletal proteins and cellular structures~\cite{Ramaswamy_nonequilibrium_2000, joanny_active_2009}, to bacteria, synthetic colloids, and animal collectives such as flocks, schools, and herds~\cite{marchetti_hydrodynamics_2013, bechinger_active_2016}. A prime example of active particles is that of bodies immersed in a solvent, which are driven into motion by gradients in solute concentration. This mechanism occurs in chemical systems as diffusiophoresis (or diffusio-osmosis)~\cite{Derjaguin_Diffusiophoresis_1961, Anderson_Colloid_1989} and in biological systems as chemotaxis~\cite{Pfeffer_Lokomotorische_1884}.

In this work we shall focus on orientational chemotaxis, but since the underlying modelling is common to many problems, we begin by casting our work within this broad context. Diffusiophoresis refers to the fluid flow past a surface that is induced by tangential pressure gradients arising from spatial variations in molecular interactions between the solute and the surface under a concentration gradient~\cite{golestanian_phoretic_2018}. It was shown in Ref.~\cite{Anderson_Colloid_1989} that in itself it can drive colloidal transport. Both the translational and the angular velocity of the particle are generically coupled to the chemical gradient~\cite{Anderson_Colloid_1989}. Catalytically active colloids are capable of generating their own solute gradients, thereby undergoing self-diffusiophoresis~\cite{Golestanian_Propulsion_2005,Golestanian_Designing_2007}. Collective interactions mediated by the solute concentration field can lead to the formation of dynamic clusters, collapses, and wave patterns~\cite{Theurkauff_Dynamic_2012,Pohl_Dynamic_2014,Liebchen_Phoretic_2017}.

Chemotaxis refers to the ability of a microorganism ({\it e.g}., a bacterium) to sense chemical gradients through its internal machinery, resulting in motion either along or against the gradient.  Bacteria can detect chemical gradients using a time-delay mechanism~\cite{de_gennes_chemotaxis_2004} or employ molecular appendages, such as pili, to spatially detect the gradient~\cite{kranz_effective_2016} and pull themselves along surfaces~\cite{Maier_Bacterial_2013}.  Keller and Segel showed that chemotactic interactions between bacteria generically drive an instability yielding the aggregation and collapse of bacterial colonies~\cite{Keller_Initiation_1970,Keller_Model_1971,Childress_Nonlinear_1981}.
Bacteria can also react to their own secreted chemicals, {\it i.e.}, self-chemotaxis, due to the coupling between the translational velocity and the chemical gradient~\cite{grima_strong-coupling_2005,Taktikos_Modeling_2011,Taktikos_Collective_2012}. This leads to slowed-down effective diffusion~\cite{grima_strong-coupling_2005,Sengupta_Dynamics_2009} or self-trapping (with slow diffusion) in the case of strong couplings~\cite{tsori_self-trapping_2004}. Depending on the microorganism, the secreted chemical may diffuse faster than the microorganism ({\it e.g.}, in the case of \textit{Dictyostelium}, neutrophils), at a comparable rate (e.g., \textit{E. coli}~\cite{Brenner_Physical_1998}), or with negligibly small diffusion coefficient ({\it e.g.}, \textit{P. aeruginosa}~\cite{zhao_psl_2013}).  In the latter case, the bacterium deposits a fixed trail of slime, which  influences the movement of other cells or its own trajectory when it crosses its path again.  

\begin{figure}
    \centering
    \includegraphics[width=.35\columnwidth,angle = 270]{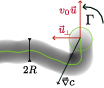}
    \caption{Sketch of a bacterium (green circle) diffusing on a substrate and leaving a trail of slime (gray ribbon) while randomly changing direction. The torque $\Gamma \propto \bm{u}_\perp \cdot \bm{\nabla} c$, which couples the angular velocity to the slime concentration gradient, enhances the rotation and can lead to a circling instability of the trajectory~\cite{kranz_effective_2016}.}
    \label{fig:instability}
\end{figure}

Recently, Kranz \textit{et al.} reported a case of orientational self-chemotaxis in {\it P. aeruginosa}, arising from the coupling between the bacterium’s angular velocity and the surrounding chemical gradient~\cite{kranz_effective_2016,gelimson_multicellular_2016} . When the trajectory of the bacterium bends, the chemical concentration increases toward the center of curvature, generating a torque that steers the particle toward regions of higher concentration, as illustrated in Fig.~\ref{fig:instability}. At sufficiently strong coupling, it is argued in Ref.~\cite{kranz_effective_2016} that this leads to an instability in which the particle rotates with a microscopic curvature radius and becomes effectively trapped. These authors analyzed the effective diffusion below the instability threshold using a theory linear in the angular velocity and validated the self-chemotaxis mechanism experimentally~\cite{gelimson_multicellular_2016}.
In this work, we build upon \cite{kranz_effective_2016} by taking into account all nonlinearities in the angular velocity, and we show that the instability disappears; self-trapping  occurs at arbitrary coupling.\\

The same equations model all the diffusiophoretic and chemotactic systems discussed above~\cite{golestanian_phoretic_2018,Pohl_Dynamic_2014,grima_strong-coupling_2005,Sengupta_Dynamics_2009,kranz_effective_2016}. In a two-dimensional setting, and in terms of the particle’s position $\bm{r}(t)$, orientation $\theta(t)$ and chemical field $c(\bm x,t)$, they read
\begin{align}
\partial_t\bm r&=v_0\bm u
+\mu\bm\nabla c(\bm r)+\sqrt{2 D_t}\,\bm \xi(t),\\
\partial_t\theta&=\mu'\bm u_\perp\cdot\bm\nabla c(\bm r)
  +\sqrt{2 D_r}\,\eta(t),
\\
\partial_tc&=D\nabla^2c-kc+ h\,\Delta_R\!\left((\bm x-\bm r)^2\right),
\label{eqc}
\end{align}
where $(\bm u(t), \bm u_\perp(t))$ is the polar basis with polar angle $\theta(t)$. Here, $v_0$ denotes the self-diffusiophoretic velocity in the case of colloids, or the particle’s self-propulsion speed in the case of bacteria, $\mu$ represents the coupling between the translational velocity and the chemical gradient (translational diffusiophoresis), $\mu'$ the coupling between the angular velocity and the chemical gradient (angular diffusiophoresis), $\bm{\xi}(t)$ is a Gaussian white noise associated with translational diffusion $D_t$, and $\eta(t)$ is a Gaussian white noise associated with rotational diffusion $D_r$. The last equation in Eq.\eqref{eqc} describes the production, degradation, and diffusion of the chemical agent, with $D$ its diffusion coefficient, $k$ the degradation rate, $h$ the chemical production rate. The function $\Delta_R$ of width $R$ describes the region over which the chemical is deposited. For a chemotactic bacterium moving on a two-dimensional substrate while depositing a trail of slime with negligible diffusion and degradation, we set $D=0$ and $k=0$. Following \cite{kranz_effective_2016}, we also assume that the noise and the chemotactic response are predominantly orientational and take $D_t=0$ and $\mu=0$. The chemical production profile is given by $\Delta_R(r) = \Theta(R^2 - r^2)/(\pi R^2)$, corresponding to a radial step function of radius~$R$. This concludes the presentation of the model we study in this work.\\

Before addressing the full nonlinear problem, let us estimate the value $\mu' = \mu'_0$ at which the angular diffusion is significantly affected by the torque, and the typical value $\mu' = \mu'_c$ around which the self-trapping discussed above might occur. We consider the angular variation produced by the torque term $\mu'\bm u_\perp\cdot\bm\nabla c$ over the time $\tau=R/v_0$ that the particle takes to move a distance comparable to the width of its trail (or equivalently, to its own size). It is given by
\begin{equation}
\label{eq:angular_variation}
\delta\theta=\mu'\bm u_\perp\cdot\bm\nabla c\times\tau\simeq\mu' R^{-1}\times\frac{ h\tau}{\pi R^2}\times\tau=\frac{\mu'h\tau^2}{\pi R^3}
\end{equation}
(see Fig.~\ref{fig:instability}).
When this deterministic $\delta\theta$ is of the same order as the stochastic one due to noise, namely $\sqrt{2D_r\tau}$, we obtain $\mu'_0$. Similarly, when $\delta\theta$ is of  order unity (in practice we take $2$), we obtain $\mu'_c$:
\begin{align}\label{thresholds}
    \mu'_0=\frac{\sqrt{2D_r}\pi R^3}{h\tau^{3/2}},\qquad
    \mu'_c=\frac{2\pi R^3}{h\tau^2}.
\end{align}
Note that this value corresponds to the threshold $\mu'_c$ of the trapping instability considered in Ref.~\cite{kranz_effective_2016}.

To simplify the theoretical analysis, we normalize  lengths by $R$, times by the bare persistence time $\tau_p=1/(2D_r)$ and we scale the concentration $c$ with $h/(2D_r R^2)$. The velocity is thus scaled with $v_0/(2D_rR)=\tau_p/\tau\equiv\varepsilon^{-1}$ and the coupling $\mu'$ by $4D_r^2R^3/h$. This rescaling amounts to setting $v_0=\varepsilon^{-1}$, $R=1$, $h=1$,  $2D_r=1$ and $\tau=\epsilon$ (together with $D=k=0)$ which we henceforth adopt. The equations of motion become
\begin{align}
\label{velocity}
\partial_t\bm r&=\varepsilon^{-1}\bm u,
\\
\partial_t\theta&=\mu'\,\bm u_\perp\cdot\bm\nabla c+\eta,
\label{eq:pourtheta}\\
\partial_t c &= \frac1\pi \Theta\!\left(1 - \left(\bm x-\bm r\right)^2\right),
\end{align}
with $\langle\eta(t)\eta(t')\rangle=\delta(t-t')$. With these normalizations, we obtain  $\mu'_c{\varepsilon}^2=2\pi$ and $\mu'_0{\varepsilon}^2=\pi\sqrt{\varepsilon}$.\\

The dynamics of the particle is actually very complex and non-Markovian, since it can cross its past trail and  interact with it many times. Similarly to Ref.~\cite{kranz_effective_2016}, our goal is not to analyze this complex diffusion, but rather to investigate the particle’s dynamics in the regime where it does not cross its past trail. The main questions we address and answer in this work are i) Can the particle experience self-trapping by interacting with its immediate trail?, ii) How is the diffusion affected before a potential trapping?\\

We first show how to arrive at a generalized Langevin equation for the angle. The torque term in the dynamical equation for $\theta(t)$ can be computed with the concentration profile $c(\bm x,t)=\int_{0}^{t} \frac1\pi \Theta(1 -(\bm x-\bm r(t'))^2)dt'$  from
\begin{align}\label{eq:tprimi}
\bm\nabla c(\bm r)
&=-\frac2\pi\int_{0}^{t} (\bm r(t)-\bm r(t'))\,\delta\!\left(1-(\bm r(t)-\bm r(t'))^2\right) \,dt',
\nonumber\\
&=
-\frac1\pi\sum_{t'_i}\frac{\bm r(t) -\bm r (t'_i)}{\left|\left(\bm r(t) -\bm r (t'_i)\right)\cdot\frac{d\bm r(t')}{dt'}|_{t'_i}\right|}
\end{align}
where the $t'_i<t$ are the times at which the particle was at a distance unity to $\bm r(t)$. These times can be computed from $\bm r(t)-\bm r(t')=\varepsilon^{-1}\!\int_{t'}^{t}\bm u(s)\,ds$. Expanding $\bm u(s)$ in a power series around $s = t$ (and collecting all terms linear in $\theta$ for later use) yields
\begin{align}\label{eq:inversion}
&\bm r(t)-\bm r(t')+\frac1\varepsilon\sum_{n=1}^\infty \frac{(t'-t)^{n+1}}{(n+1)!}\, \theta^{(n)}\bm u_\perp
=\frac{t-t'}\varepsilon\bm u
\nonumber\\
&-\frac{\dot\theta^2}{6}\frac{(t-t')^3}\varepsilon\bm u
+\frac{1}{24}\frac{(t-t')^4}\varepsilon\left(\dot{\theta}^3\bm u_\perp+3\dot{\theta}\ddot{\theta}\bm u\right)
+\ldots,
\end{align}
where the ellipsis refer to $O((t-t')^5)$ terms and where $\theta$, $\bm u$ and $\bm u_\perp$ are evaluated at time $t$. Since, we neglect trail-crossings, there is only one root $t'_i=t'_1$ of the equation $|\bm r(t)-\bm r(t'_i)|=1$. Formally we can thus express $t-t'_1$ as a series in powers  $\varepsilon$ with coefficients that are polynomials in the derivatives of $\theta$ at time $t$. We find that $t'_1=t-\varepsilon-\frac{1}{24}\dot{\theta}^2\varepsilon^3+\frac1{24}\dot\theta\ddot\theta\varepsilon^4+O(\varepsilon^5)$. Indeed, if the particle had a straight trajectory, it would take a time $R/v_0=\varepsilon$ to run over a distance equal to its size $R$, but if this trajectory weakly deforms, there are curvature corrections. We thus obtain an effective equation for the angular velocity alone:
\begin{align}\label{eq:eqnonlin}
\dot\theta\!=\!
\frac{\mu'\!\varepsilon^2}\pi\!\left[\sum_{n=1}^\infty\,\frac{(-\varepsilon)^{n-1}}{(n+1)!}\theta^{(n)}\!+\!\frac{\dot{\theta}^3}{16}\varepsilon^2\! -\frac7{60}\dot{\theta}^2\ddot{\theta}\varepsilon^3\!+\! O(\varepsilon^4)\right]\!+\eta.
\end{align}
Even though we have truncated this expansion to $O(\varepsilon^4)$, we have deliberately kept for each order in $\varepsilon$ the coefficient that is linear in $\theta$. Interestingly, the sum over $n$ in the above equation  (the linear terms in $\theta$) can be resummed into $\int_0^{\varepsilon} dt'\, (\varepsilon - t') \dot{\theta}(t - t')$. The approximation carried out in  Ref.~\cite{kranz_effective_2016} amounts to keeping only that contribution in Eq.~\eqref{eq:eqnonlin}. There are no other linear contributions in $\theta$ appearing in Eq.~\eqref{eq:eqnonlin}. At this linear level, the analysis~\cite{kranz_effective_2016} shows the existence of a threshold for a trapping instability at $\mu'\varepsilon^2=2\pi$.

\begin{figure}
    \centering
    \includegraphics[width=.9\columnwidth]{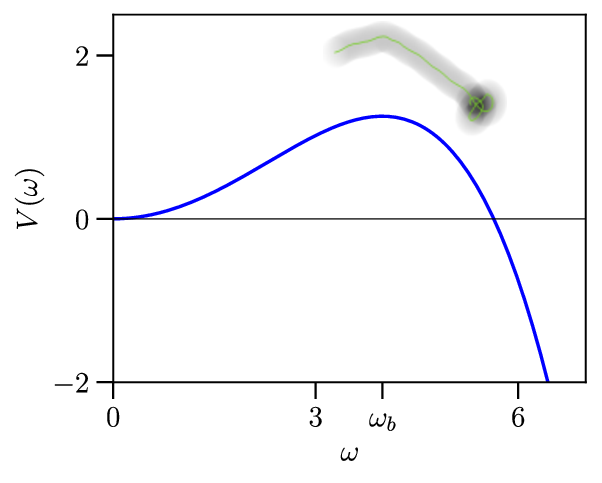}
    \caption{Fourth-order polynomial potential $V(\omega)$ governing the diffusion of the particle’s angular velocity in the regime $\mu'\lesssim\mu'_c$ and $\varepsilon\ll1$, shown for $\Delta\hat\mu'=0.3$ and $\varepsilon=0.1$. The trapping instability corresponds to the passage of the potential barrier. Inset: Trapping instability observed in our numerical simulation for $\mu'/\mu'_c=0.49$ and $\varepsilon=0.1$. The green line shows the particle’s trajectory, and the grey levels indicate the deposited slime.}
    \label{fig:trapping_potential}
\end{figure}

In order to explore the effect of keeping the nonlinearities, we find it convenient to define
\begin{align}
    \hat{\mu}'=\mu'\varepsilon^2,
\end{align}
and to work at $\varepsilon\to 0$ while keeping $\hat{\mu}'$
of order unity. The effective dynamics of the angular velocity $\dot{\theta}$ reads
\begin{align}
    &\left(1-\frac{\hat{\mu}'}{\hat{\mu}_c'}\right)\dot\theta=
    \frac{\hat{\mu}'\varepsilon}\pi\bigg[
    -\frac{\ddot\theta}{6}
    +\left(\frac{\dot\theta^3}{16}+\frac{\theta^{(3)}}{24}\right)\varepsilon
    \nonumber\\
    &\hspace{1.5cm}-\left(
    \frac{7\dot\theta^2\ddot\theta}{60}+\frac{\theta^{(4)}}{120}
    \right)\varepsilon^2+O\left(\varepsilon^3\right)
    \bigg]+\eta(t),
\end{align}
with $\hat{\mu}'_c = 2\pi$. 
When $\hat\mu'$ is close to the characteristic coupling $\hat{\mu}'_c$, this stochastic equation can be mapped onto a standard Langevin equation for $\dot{\theta}$ by using the following scaling variables:
\begin{align}
    \Delta\hat{\mu}'&=\varepsilon^{-\frac12}\!\left(1-\frac{\hat{\mu}'}{\hat{\mu}'_c}\right),\quad
    \bar t=\varepsilon^{-\frac12}t,\quad
    \\
    \bar\theta(\bar t)&=\varepsilon^{\frac14}\theta(t),\quad
    \bar\eta(\bar t)=\varepsilon^{\frac14}\eta(t).
\end{align}
In the limit $\varepsilon\to0$, we obtain an overdamped Langevin equation in which the angular velocity 
 $\bar\omega(\bar t)=d\bar{\theta}/d\bar{t}$ effectively evolves in the quartic potential $V(\bar\omega)=\frac12\Delta\hat{\mu}'\,\bar\omega^2-\bar\omega^4/(64\pi)$,
\begin{align}\label{eq:Langevin}
    0=-\frac{\hat{\mu}'}{6\pi}\dot{\bar\omega}-V'(\bar\omega)+\bar\eta(\bar t)+O\left(\varepsilon\right),
\end{align}
with $\langle\bar\eta(\bar t)\bar\eta(\bar t')\rangle=\delta(\bar t-\bar t')$. This potential is destabilizing not only for $\hat{\mu}'  \ge \hat{\mu}'_c$, as identified in Ref.~\cite{kranz_effective_2016}, but also for $\hat{\mu}'< \hat{\mu}'_c$. Indeed, the effective barrier can be overcome through an activated process,
corresponding to the trapping of the bacterium (see Fig.~\ref{fig:trapping_potential}).
The barrier height being small as $\sim\!\Delta{\hat \mu}'^2$, the scaled crossing time $\bar T_\text{fp}(\bar\omega_0)$---the time required to reach the barrier at $\bar\omega=\bar\omega_b$ for the first time starting from $\bar\omega=\bar\omega_0$---cannot be estimated using Kramers' formula. Instead, it can be obtained from the following first-passage differential equation~\cite{van_kampen_stochastic_2007}:
\begin{align}
\label{eq:first_passage_time}
-\frac{6\pi}{\hat{\mu}'}V'(\bar\omega_0)
\bar T_\text{fp}'(\bar\omega_0)
+\frac 12\left(\frac{6\pi}{\hat{\mu}'}\right)^2
\bar T_\text{fp}''(\bar\omega_0)=-1,
\end{align}
with boundary condition $\bar T_\text{fp}(\pm\bar\omega_b)=0$. Solving this equation perturbatively for $\bar\omega_b\to0$ (see SM~\cite{noauthor_notitle_nodate}), we obtain for the mean trapping time starting from $\bar\omega_0=0$, $T=T_\text{fp}(0)$, the linear dependance close to the characteristic coupling $\hat\mu'_c$:
\begin{align}
\label{eq:T_near_threshold}
T\simeq\frac{16\pi}9 \left(1-\frac{\mu'}{\mu'_c}\right),
\qquad(\mu'<\mu'_c).
\end{align}

The other analytic limit is obtained for $\varepsilon\to0$ while keeping $\hat{\mu}'$ of order unity, yielding
\begin{align}
    \left(1-\frac{\hat\mu'}{\hat\mu'_c}\right)\dot\theta=\eta(t)+O(\varepsilon).
\end{align}

This leads to an effective angular diffusivity $D_\theta$ and an effective translational diffusivity $D_\text{tr}$ given by~\cite{romanczuk_active_2012}:
\begin{align}
\label{Diffusion}
    D_\theta(\mu')=\frac12\left(1-\frac{\mu'}{\mu'_c}\right)^{-2},
    \quad
    D_\text{tr}(\mu')=\frac1{2\varepsilon^2 D_\theta(\mu')}.
\end{align}

In order to probe the range of validity of these asymptotic analytical predictions, we perform extensive Brownian dynamics simulations, for $\varepsilon=0.1$, as follows. From the stored discretized trajectory $\bm{r}(t)$, obtained with a time step $\Delta t = 10^{-3}$, we identify using Eq.~\eqref{eq:tprimi} the times $t'_i$ that contribute to $\bm{\nabla} c(\bm{r}(t))$. The orientation $\theta(t)$ is then updated using an Euler scheme, including the stochastic noise term, followed by the update of the particle position $\bm{r}(t)$.

\begin{figure}
    \centering
    \includegraphics[width=.9\columnwidth]{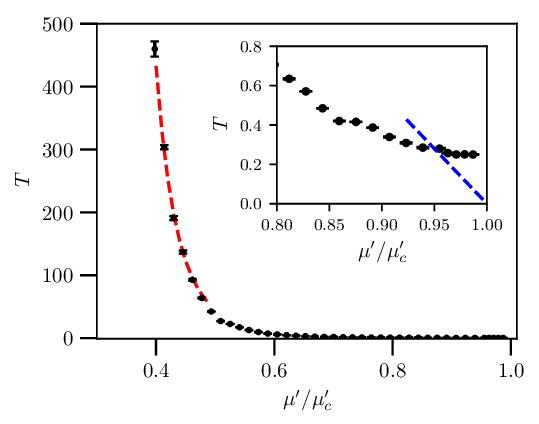}
    \caption{Mean trapping time $T$ for $\mu'<\mu'_c$, obtained from the statistics of our numerical simulations. The blue dashed line in the inset shows the analytical prediction from the mean first-passage time corresponding to Eq.~\eqref{eq:T_near_threshold}. The red dashed line in the main plot represents a one parameter fit to the trapping time estimated with Eq.~\eqref{eq:T_asymptotic}.}
    \label{fig:trapping_time}
\end{figure}

We observe numerically that trapping typically occurs when the radius of curvature $r=1/(\omega \varepsilon)$ is of order unity; in practice we take $r \simeq 0.4$. To determine the mean self-trapping time~$T$, we erase the tail of the trail after a short duration, which allows us to avoid most of the crossings. The remaining ones are dealt with by using a statistical method explained in the SM \cite{noauthor_notitle_nodate}. In Fig.~\ref{fig:trapping_time}, we show that the particle does self-trap below the characteristic coupling ${\mu}' = {\mu}'_c$. We find that $T$ is much smaller than the bare persistence time near $\mu'_c$, increases only slowly for $\mu' \le \mu'_c$, and then rises sharply. Note that for $\mu' \to \mu'_c$ the agreement with the theoretical prediction~\eqref{eq:T_near_threshold}  is only trend-wise (see inset of Fig.~\ref{fig:trapping_time}), since in the theory the trapping time $T$ is defined by barrier crossing, while in the simulations it corresponds to reaching a threshold angular velocity.

From the numerics in Fig.~\ref{fig:trapping_time}, a natural question is whether the mean trapping time diverges at a finite value  or whether the observed growth actually marks the beginning of a divergence for $\mu' \to 0$. To investigate whether the particle can self-trap for arbitrary values of $\mu' > 0$, we consider a scenario in which the noise increases the trajectory’s curvature, causing the particle to follow a circular path with constant angular velocity $\dot\theta=\omega$ and microscopic radius $r =(\omega \varepsilon)^{-1}= 1$ over a total polar angle $\varphi\gg1$. This repeated circling amplifies $\bm\nabla c$ and therefore enhances the torque term $\Gamma = \mu'\bm\nabla c \cdot \bm u_\perp$, even for small values of $\mu'$. The trapping probability $P_\text{traj}$ is then determined by the following equations:
\begin{align}\label{eq:Gammavarphi}
  &\Gamma(\varphi)=
  \frac{\mu'\varepsilon}{\pi \sqrt{3}}\left(
  \lfloor \frac{\varphi}{2\pi} +\frac16\rfloor 
  + \lfloor \frac{\varphi}{2\pi}+\frac56 \rfloor \right),
  \\\label{eq:varphistar}
  &\Gamma(\varphi^\star)\varepsilon=1,
  \\\label{eq:Ptraj}
  &P_\text{traj}\sim\exp\left(-\int_0^{\varphi^\star}\! \frac{d\varphi}\omega\, \frac{\eta^2}2\right)\sim T^{-1}.
\end{align}
In the SM~\cite{noauthor_notitle_nodate} we show that for an angle $\varphi$ the torque is given by Eq.~\eqref{eq:Gammavarphi}. With Eq.~\eqref{eq:varphistar} we obtain the typical trapping angle $\varphi^\star$ as discussed in Eq.~\eqref{eq:angular_variation}. Equation~\eqref{eq:Ptraj} with $\eta = \omega - \Gamma$ (see Eq.~\eqref{eq:pourtheta}), gives the Onsager-Machlup probability that a noise realization $\eta(t)$ produces the circling trajectory under consideration and the associated scaling of the trapping time.
For $\mu'\to0$, we obtain $\varphi^\star\sim1/(\mu'\varepsilon^2)$ (see SM~\cite{noauthor_notitle_nodate}) yielding the prediction of a strongly diverging trapping time,
\begin{align}
\label{eq:T_asymptotic}
    T\sim\exp\left(\frac\pi{4\varepsilon\sqrt{3}}\,\frac{\mu'_c}{\mu'}\right),
\end{align}
suggesting the absence of a threshold for the trapping instability. Note that this formula captures well the divergence of $T$ seen in Fig.~\ref{fig:trapping_time} even though $\mu'/\mu'_c\simeq0.4$ is not small.

\begin{figure}
    \centering
    \includegraphics[width=.9\columnwidth]{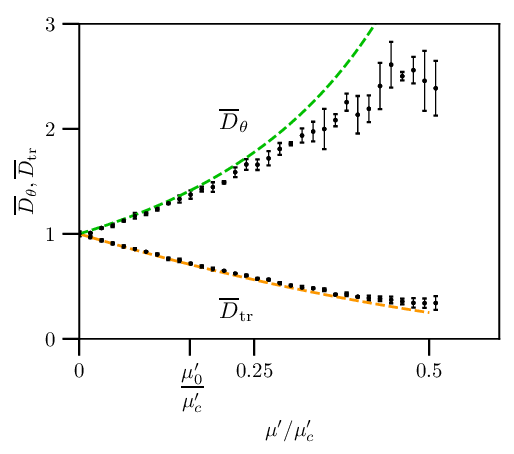}
    \caption{Rotational ($\overline{D}_\theta$) and translational ($\overline{D}_\text{tr}$) diffusion coefficients of the particle for $\varepsilon=0.1$, each normalized by their values at $\mu'=0$. The dashed lines show the analytical predictions in the limit $\varepsilon \to 0$ (Eq.~\eqref{Diffusion}). The graphs end when the trapping time is too short to allow a well-defined diffusive regime. The constant $\mu'_0$ is defined in Eq.~\eqref{thresholds}.
    }
    \label{fig:Diffusion_coefficients}
\end{figure}

When the trapping time $T\gg1$, it is physically meaningful to consider the particle’s diffusion coefficients before trapping occurs. Fig.~\ref{fig:Diffusion_coefficients} shows the angular and translational diffusion coefficients measured numerically. Note that they are well fitted for $\mu'\ll1$ by the formulas given in Eq.~\eqref{Diffusion} obtained in the limit $\varepsilon\to0$. The error bars increase when $T$ is less than a few decades, as statistical sampling deteriorates.\\

In summary, we have explored the chemotactic bacterium model proposed in Ref.~\cite{kranz_effective_2016} and the related trapping instability of the bacterium's trajectory. We found that there is no threshold in the chemical coupling allowing to avoid this instability, owing to nonlinear mechanisms. It is interesting to note that in the weak coupling $\mu'\to 0$ limit, the angular diffusion coefficient matches that of \cite{kranz_effective_2016} up to $\mu'\simeq 0.2$, thus confirming that their model is physically relevant for their {\it P. aeruginosa} system in that regime~\cite{gelimson_multicellular_2016}. It would be interesting to find out whether, in the trapping regime, the particle undergoes translational diffusion or whether it remains stuck at a given location in space. Does the rotational diffusion affect the scenario discussed in Refs.~\cite{tsori_self-trapping_2004, grima_strong-coupling_2005}? In our work, we have neglected translational noise ($D_t=0$), but it is for us an open question whether restoring a nonzero $D_t$ could help the particle escape the instability, with or without a threshold in $D_t$ (a similar question arises for $D$ that controls the slime diffusion). In the same vein, we have limited our analysis to self-trapping occurring at the tip of the trajectory, although both diffusion and trapping should be affected by the whole imprinted trail left by the bacterium in the past, which raises another set of questions. Finally, in assembling many of the bacteria, it would be very interesting to find out how individual trails affect collective behavior~\cite{mahdisoltani_nonequilibrium_2021}.
\\

We thank Ramin Golestanian for several useful discussions. FvW acknowledges the support of the ANR grant THEMA.

\bibliographystyle{apsrev4-2}
\bibliography{Trail}

@article{golestanian_propulsion_2005,
	title = {Propulsion of a {Molecular} {Machine} by {Asymmetric} {Distribution} of {Reaction} {Products}},
	volume = {94},
	url = {https://link.aps.org/doi/10.1103/PhysRevLett.94.220801},
	doi = {10.1103/PhysRevLett.94.220801},
	number = {22},
	urldate = {2024-09-26},
	journal = {Phys. Rev. Lett.},
	author = {Golestanian, Ramin and Liverpool, Tanniemola B. and Ajdari, Armand},
	month = jun,
	year = {2005},
	note = {Publisher: American Physical Society},
	pages = {220801},
}

@article{golestanian_designing_2007,
	title = {Designing phoretic micro- and nano-swimmers},
	volume = {9},
	issn = {1367-2630},
	url = {https://iopscience.iop.org/article/10.1088/1367-2630/9/5/126},
	doi = {10.1088/1367-2630/9/5/126},
	language = {en},
	number = {5},
	urldate = {2024-10-01},
	journal = {New J. Phys.},
	author = {Golestanian, R and Liverpool, T B and Ajdari, A},
	month = may,
	year = {2007},
	pages = {126--126},
}

@article{kranz_effective_2016,
	title = {Effective {Dynamics} of {Microorganisms} {That} {Interact} with {Their} {Own} {Trail}},
	volume = {117},
	copyright = {http://link.aps.org/licenses/aps-default-license},
	issn = {0031-9007, 1079-7114},
	url = {https://link.aps.org/doi/10.1103/PhysRevLett.117.038101},
	doi = {10.1103/PhysRevLett.117.038101},
	language = {en},
	number = {3},
	urldate = {2024-10-01},
	journal = {Phys. Rev. Lett.},
	author = {Kranz, W. Till and Gelimson, Anatolij and Zhao, Kun and Wong, Gerard C. L. and Golestanian, Ramin},
	month = jul,
	year = {2016},
	pages = {038101},
}

@article{keller_model_1971,
	title = {Model for chemotaxis},
	volume = {30},
	copyright = {https://www.elsevier.com/tdm/userlicense/1.0/},
	issn = {00225193},
	url = {https://linkinghub.elsevier.com/retrieve/pii/0022519371900506},
	doi = {10.1016/0022-5193(71)90050-6},
	number = {2},
	urldate = {2024-10-01},
	journal = {Journal of Theoretical Biology},
	author = {Keller, Evelyn F. and Segel, Lee A.},
	month = feb,
	year = {1971},
	pages = {225--234},
}

@article{theurkauff_dynamic_2012,
	title = {Dynamic {Clustering} in {Active} {Colloidal} {Suspensions} with {Chemical} {Signaling}},
	volume = {108},
	copyright = {http://link.aps.org/licenses/aps-default-license},
	issn = {0031-9007, 1079-7114},
	url = {https://link.aps.org/doi/10.1103/PhysRevLett.108.268303},
	doi = {10.1103/PhysRevLett.108.268303},
	language = {en},
	number = {26},
	urldate = {2024-10-02},
	journal = {Phys. Rev. Lett.},
	author = {Theurkauff, I. and Cottin-Bizonne, C. and Palacci, J. and Ybert, C. and Bocquet, L.},
	month = jun,
	year = {2012},
	pages = {268303},
}

@article{pohl_dynamic_2014,
	title = {Dynamic {Clustering} and {Chemotactic} {Collapse} of {Self}-{Phoretic} {Active} {Particles}},
	volume = {112},
	copyright = {http://link.aps.org/licenses/aps-default-license},
	issn = {0031-9007, 1079-7114},
	url = {https://link.aps.org/doi/10.1103/PhysRevLett.112.238303},
	doi = {10.1103/PhysRevLett.112.238303},
	number = {23},
	urldate = {2024-10-03},
	journal = {Phys. Rev. Lett.},
	author = {Pohl, Oliver and Stark, Holger},
	month = jun,
	year = {2014},
	pages = {238303},
}

@article{brenner_physical_1998,
	title = {Physical {Mechanisms} for {Chemotactic} {Pattern} {Formation} by {Bacteria}},
	volume = {74},
	copyright = {https://www.elsevier.com/tdm/userlicense/1.0/},
	issn = {00063495},
	url = {https://linkinghub.elsevier.com/retrieve/pii/S0006349598778804},
	doi = {10.1016/S0006-3495(98)77880-4},
	language = {en},
	number = {4},
	urldate = {2024-10-07},
	journal = {Biophysical Journal},
	author = {Brenner, Michael P. and Levitov, Leonid S. and Budrene, Elena O.},
	month = apr,
	year = {1998},
	pages = {1677--1693},
}

@article{childress_nonlinear_1981,
	title = {Nonlinear aspects of chemotaxis},
	volume = {56},
	copyright = {https://www.elsevier.com/tdm/userlicense/1.0/},
	issn = {00255564},
	url = {https://linkinghub.elsevier.com/retrieve/pii/0025556481900559},
	doi = {10.1016/0025-5564(81)90055-9},
	number = {3-4},
	urldate = {2024-10-07},
	journal = {Mathematical Biosciences},
	author = {Childress, S. and Percus, J.K.},
	month = oct,
	year = {1981},
	pages = {217--237},
}

@article{taktikos_collective_2012,
	title = {Collective dynamics of model microorganisms with chemotactic signaling},
	volume = {85},
	copyright = {http://link.aps.org/licenses/aps-default-license},
	issn = {1539-3755, 1550-2376},
	url = {https://link.aps.org/doi/10.1103/PhysRevE.85.051901},
	doi = {10.1103/PhysRevE.85.051901},
	language = {en},
	number = {5},
	urldate = {2024-10-07},
	journal = {Phys. Rev. E},
	author = {Taktikos, Johannes and Zaburdaev, Vasily and Stark, Holger},
	month = may,
	year = {2012},
	pages = {051901},
}

@article{keller_initiation_1970,
	title = {Initiation of slime mold aggregation viewed as an instability},
	volume = {26},
	copyright = {https://www.elsevier.com/tdm/userlicense/1.0/},
	issn = {00225193},
	url = {https://linkinghub.elsevier.com/retrieve/pii/0022519370900925},
	doi = {10.1016/0022-5193(70)90092-5},
	number = {3},
	urldate = {2024-10-08},
	journal = {Journal of Theoretical Biology},
	author = {Keller, Evelyn F. and Segel, Lee A.},
	month = mar,
	year = {1970},
	pages = {399--415},
}

@article{gelimson_multicellular_2016,
	title = {Multicellular {Self}-{Organization} of {P}. aeruginosa due to {Interactions} with {Secreted} {Trails}},
	volume = {117},
	journal = {Phys. Rev. Lett.},
	author = {Gelimson, Anatolij and Zhao, Kun and Lee, Calvin K and Kranz, W Till and Wong, Gerard C L and Golestanian, Ramin},
	year = {2016},
	pages = {178102},
}

@article{taktikos_modeling_2011,
	title = {Modeling a self-propelled autochemotactic walker},
	volume = {84},
	copyright = {http://link.aps.org/licenses/aps-default-license},
	issn = {1539-3755, 1550-2376},
	url = {https://link.aps.org/doi/10.1103/PhysRevE.84.041924},
	doi = {10.1103/PhysRevE.84.041924},
	language = {en},
	number = {4},
	urldate = {2025-05-08},
	journal = {Phys. Rev. E},
	author = {Taktikos, Johannes and Zaburdaev, Vasily and Stark, Holger},
	month = oct,
	year = {2011},
	pages = {041924},
}

@article{grima_strong-coupling_2005,
	title = {Strong-{Coupling} {Dynamics} of a {Multicellular} {Chemotactic} {System}},
	volume = {95},
	copyright = {http://link.aps.org/licenses/aps-default-license},
	issn = {0031-9007, 1079-7114},
	url = {https://link.aps.org/doi/10.1103/PhysRevLett.95.128103},
	doi = {10.1103/PhysRevLett.95.128103},
	language = {en},
	number = {12},
	urldate = {2025-05-08},
	journal = {Phys. Rev. Lett.},
	author = {Grima, R.},
	month = sep,
	year = {2005},
	pages = {128103},
}

@article{sengupta_dynamics_2009,
	title = {Dynamics of a microorganism moving by chemotaxis in its own secretion},
	volume = {80},
	copyright = {http://link.aps.org/licenses/aps-default-license},
	issn = {1539-3755, 1550-2376},
	url = {https://link.aps.org/doi/10.1103/PhysRevE.80.031122},
	doi = {10.1103/PhysRevE.80.031122},
	language = {en},
	number = {3},
	urldate = {2025-05-20},
	journal = {Phys. Rev. E},
	author = {Sengupta, Ankush and Van Teeffelen, Sven and Löwen, Hartmut},
	month = sep,
	year = {2009},
	pages = {031122},
}

@article{tsori_self-trapping_2004,
	title = {Self-trapping of a single bacterium in its own chemoattractant},
	volume = {66},
	issn = {0295-5075, 1286-4854},
	url = {https://iopscience.iop.org/article/10.1209/epl/i2003-10237-5},
	doi = {10.1209/epl/i2003-10237-5},
	language = {en},
	number = {4},
	urldate = {2025-10-20},
	journal = {Europhys. Lett.},
	author = {Tsori, Y and Gennes, P.-G. De},
	month = may,
	year = {2004},
	pages = {599--602},
}

@article{pfeffer_lokomotorische_1884,
	title = {Lokomotorische {Richtungsbewegungen} durch chemische {Reize}},
	volume = {1},
	journal = {Unters. Bot. Inst. Tübingen},
	author = {Pfeffer, Wilhelm F.},
	year = {1884},
	pages = {363--370},
}

@article{anderson_colloid_1989,
	title = {Colloid {Transport} by {Interfacial} {Forces}},
	volume = {21},
	journal = {Ann. Rev. Fluid Mech.},
	author = {Anderson, J L},
	year = {1989},
	pages = {61},
}

@article{marchetti_hydrodynamics_2013,
	title = {Hydrodynamics of soft active matter},
	volume = {85},
	copyright = {http://link.aps.org/licenses/aps-default-license},
	issn = {0034-6861, 1539-0756},
	url = {https://link.aps.org/doi/10.1103/RevModPhys.85.1143},
	doi = {10.1103/RevModPhys.85.1143},
	number = {3},
	urldate = {2025-10-23},
	journal = {Rev. Mod. Phys.},
	author = {Marchetti, M. C. and Joanny, J. F. and Ramaswamy, S. and Liverpool, T. B. and Prost, J. and Rao, Madan and Simha, R. Aditi},
	month = jul,
	year = {2013},
	pages = {1143--1189},
}

@article{derjaguin_diffusiophoresis_1961,
	title = {Diffusiophoresis in electrolyte solutions and its role in the mechanism of film formation from rubber latexes by the method of ionic deposition},
	volume = {23},
	journal = {Kolloidn. Zh.},
	author = {Derjaguin, B. V. and Dukhin, S. S. and Korotkova, A. A.},
	year = {1961},
	pages = {53},
}

@incollection{golestanian_phoretic_2018,
	series = {Lecture {Notes} of the {Les} {Houches} {Summer} {School}},
	title = {Phoretic {Active} {Matter}},
	volume = {112},
	isbn = {978-0-19-285831-3},
	url = {https://doi.org/10.1093/oso/9780192858313.003.0008},
	booktitle = {Active {Matter} and {Nonequilibrium} {Statistical} {Physics}},
	publisher = {Oxford University Press},
	author = {Golestanian, Ramin},
	year = {2018},
}

@article{de_gennes_chemotaxis_2004,
	title = {Chemotaxis: the role of internal delays},
	volume = {33},
	copyright = {http://www.springer.com/tdm},
	issn = {0175-7571, 1432-1017},
	shorttitle = {Chemotaxis},
	url = {http://link.springer.com/10.1007/s00249-004-0426-z},
	doi = {10.1007/s00249-004-0426-z},
	number = {8},
	urldate = {2025-10-23},
	journal = {Eur Biophys J},
	author = {De Gennes, P.-G.},
	month = dec,
	year = {2004},
	pages = {691--693},
}

@article{maier_bacterial_2013,
	title = {The bacterial type {IV} pilus system – a tunable molecular motor},
	volume = {9},
	issn = {1744-683X, 1744-6848},
	url = {https://xlink.rsc.org/?DOI=c3sm50546d},
	doi = {10.1039/c3sm50546d},
	number = {24},
	urldate = {2025-10-23},
	journal = {Soft Matter},
	author = {Maier, Berenike},
	year = {2013},
	pages = {5667},
}

@article{zhao_psl_2013,
	title = {Psl trails guide exploration and microcolony formation in {Pseudomonas} aeruginosa biofilms},
	volume = {497},
	issn = {1476-4687},
	url = {https://doi.org/10.1038/nature12155},
	doi = {10.1038/nature12155},
	number = {7449},
	journal = {Nature},
	author = {Zhao, Kun and Tseng, Boo Shan and Beckerman, Bernard and Jin, Fan and Gibiansky, Maxsim L. and Harrison, Joe J. and Luijten, Erik and Parsek, Matthew R. and Wong, Gerard C. L.},
	month = may,
	year = {2013},
	pages = {388--391},
}

@book{van_kampen_stochastic_2007,
	address = {Amsterdam},
	edition = {3rd},
	title = {Stochastic {Processes} in {Physics} and {Chemistry}},
	isbn = {978-0-444-52965-7},
	publisher = {North-Holland},
	author = {van Kampen, N. G.},
	year = {2007},
}

@article{bechinger_active_2016,
	title = {Active {Particles} in {Complex} and {Crowded} {Environments}},
	volume = {88},
	copyright = {http://link.aps.org/licenses/aps-default-license},
	issn = {0034-6861, 1539-0756},
	url = {https://link.aps.org/doi/10.1103/RevModPhys.88.045006},
	doi = {10.1103/RevModPhys.88.045006},
	language = {en},
	number = {4},
	urldate = {2025-10-29},
	journal = {Rev. Mod. Phys.},
	author = {Bechinger, Clemens and Di Leonardo, Roberto and Löwen, Hartmut and Reichhardt, Charles and Volpe, Giorgio and Volpe, Giovanni},
	month = nov,
	year = {2016},
	pages = {045006},
}

@article{ramaswamy_nonequilibrium_2000,
	title = {Nonequilibrium {Fluctuations}, {Traveling} {Waves}, and {Instabilities} in {Active} {Membranes}},
	volume = {84},
	copyright = {http://link.aps.org/licenses/aps-default-license},
	issn = {0031-9007, 1079-7114},
	url = {https://link.aps.org/doi/10.1103/PhysRevLett.84.3494},
	doi = {10.1103/PhysRevLett.84.3494},
	language = {en},
	number = {15},
	urldate = {2025-10-29},
	journal = {Phys. Rev. Lett.},
	author = {Ramaswamy, Sriram and Toner, John and Prost, Jacques},
	month = apr,
	year = {2000},
	pages = {3494--3497},
}

@article{joanny_active_2009,
	title = {Active gels as a description of the actin‐myosin cytoskeleton},
	volume = {3},
	issn = {1955-2068},
	url = {https://www.tandfonline.com/doi/full/10.2976/1.3054712},
	doi = {10.2976/1.3054712},
	language = {en},
	number = {2},
	urldate = {2025-10-29},
	journal = {HFSP Journal},
	author = {Joanny, Jean‐François and Prost, Jacques},
	month = apr,
	year = {2009},
	pages = {94--104},
}

@article{noauthor_notitle_nodate,
	journal = {See Supplementary Material},
}

@article{liebchen_phoretic_2017,
	title = {Phoretic {Interactions} {Generically} {Induce} {Dynamic} {Clusters} and {Wave} {Patterns} in {Active} {Colloids}},
	volume = {118},
	copyright = {http://link.aps.org/licenses/aps-default-license},
	issn = {0031-9007, 1079-7114},
	url = {http://link.aps.org/doi/10.1103/PhysRevLett.118.268001},
	doi = {10.1103/PhysRevLett.118.268001},
	language = {en},
	number = {26},
	urldate = {2025-10-31},
	journal = {Phys. Rev. Lett.},
	author = {Liebchen, Benno and Marenduzzo, Davide and Cates, Michael E.},
	month = jun,
	year = {2017},
	pages = {268001},
}

@article{mahdisoltani_nonequilibrium_2021,
	title = {Nonequilibrium polarity-induced chemotaxis: {Emergent} {Galilean} symmetry and exact scaling exponents},
	volume = {3},
	issn = {2643-1564},
	shorttitle = {Nonequilibrium polarity-induced chemotaxis},
	url = {https://link.aps.org/doi/10.1103/PhysRevResearch.3.013100},
	doi = {10.1103/PhysRevResearch.3.013100},
	language = {en},
	number = {1},
	urldate = {2025-11-27},
	journal = {Phys. Rev. Research},
	author = {Mahdisoltani, Saeed and Zinati, Riccardo Ben Alì and Duclut, Charlie and Gambassi, Andrea and Golestanian, Ramin},
	month = jan,
	year = {2021},
	pages = {013100},
}

@article{romanczuk_active_2012,
	title = {Active {Brownian} particles: {From} individual to collective stochastic dynamics},
	volume = {202},
	copyright = {http://www.springer.com/tdm},
	issn = {1951-6355, 1951-6401},
	shorttitle = {Active {Brownian} particles},
	url = {http://link.springer.com/10.1140/epjst/e2012-01529-y},
	doi = {10.1140/epjst/e2012-01529-y},
	language = {en},
	number = {1},
	urldate = {2025-12-09},
	journal = {Eur. Phys. J. Spec. Top.},
	author = {Romanczuk, P. and Bär, M. and Ebeling, W. and Lindner, B. and Schimansky-Geier, L.},
	month = mar,
	year = {2012},
	pages = {1--162},
}

\end{document}